\documentclass[aps,10pt,prb,twocolumn,amssymb,amsmath,amsfonts,showpacs,floatfix,citeautoscript,preprintnumbers,reprint,footinbib,superscriptaddress,a4paper]{revtex4-1}
\usepackage{times}
\usepackage{graphicx}

\usepackage{bm}
\usepackage{xcolor}
\newcommand{\ket}[1]{\vert{#1}\rangle}
\newcommand{\meanvalue}[1]{\langle{#1}\rangle}
\newcommand{\ups}{\uparrow}
\newcommand{\downs}{\downarrow}
\newcommand{\dimer}{\includegraphics[width=4mm,clip]{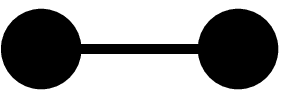}}
\newcommand{\spinon}{\includegraphics[width=2.2cm,clip]{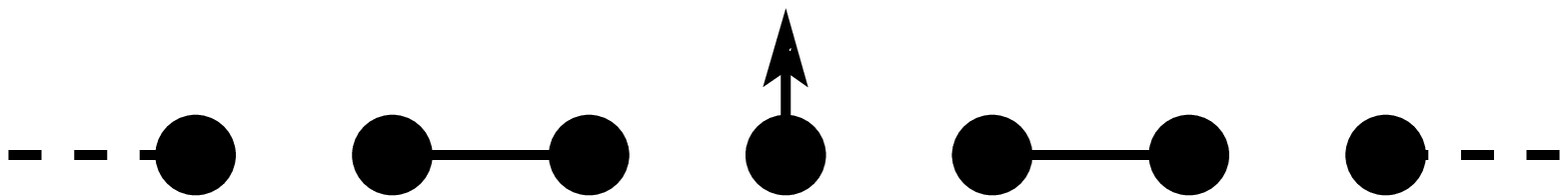}}

\begin{document}
\title{Generation of chiral solitons in antiferromagnetic chains by a quantum quench}

\author{Barbara Bravo}
\affiliation{
Facultad de Ciencias Exactas Ingenier{\'\i}a and Agrimensura, Universidad Nacional de Rosario and Instituto de
F\'{\i}sica Rosario, Bv. 27 de Febrero 210 bis, 2000 Rosario,
Argentina.}
\author{Ariel Dobry}
\affiliation{
Facultad de Ciencias Exactas Ingenier{\'\i}a and Agrimensura, Universidad Nacional de Rosario and Instituto de
F\'{\i}sica Rosario, Bv. 27 de Febrero 210 bis, 2000 Rosario,
Argentina.}
\author{Diego Mastrogiuseppe}
\affiliation{Department of Physics and Astronomy, and Nanoscale
and Quantum Phenomena Institute, \\ Ohio University, Athens, Ohio
45701--2979}
\affiliation{Dahlem Center for Complex Quantum Systems and Fachbereich Physik,
Freie Universit\"at Berlin, 14195 Berlin, Germany}
\author{Claudio Gazza}
\affiliation{
Facultad de Ciencias Exactas Ingenier{\'\i}a and Agrimensura, Universidad Nacional de Rosario and Instituto de
F\'{\i}sica Rosario, Bv. 27 de Febrero 210 bis, 2000 Rosario,
Argentina.}

\date{\today}

%%%%%%%%%%%%%%%%%%%%%%%%%%%%%%%%%%%%%%%%%%%%%%%%%%%%%%%%%

\begin{abstract}

We analyze the time evolution of a magnetic excitation in a spin-$\frac12$ antiferromagnetic Heisenberg chain after a quantum quench.
By a proper modulation of the magnetic exchange coupling, we prepare a static soliton of total spin $\frac{1}{2}$ as an initial spin state.
Using bosonization and a numerical time dependent density matrix renormalization group algorithm,
we show that the initial excitation evolves to a state composed of two counter-propagating
chiral states, which interfere to yield $\meanvalue{S^z} = \frac14$ for each mode.
We find that these dynamically generated states remain considerably stable as time evolution is carried out.
We propose spin-Peierls materials and ultracold-atom systems as suitable experimental scenarios in which to conduct and observe this mechanism.
\end{abstract}


\maketitle
\section{INTRODUCTION}
Thinking about classical nonlinear physics, solitons are peculiar solutions which can be 
characterized by constant velocity and shape. Recently, W\"ollert and Honecker \cite{AWAH_12} pursuing 
the understanding of the extension of the soliton concept to the quantum regime, chose the easy-axis ferromagnetic $XXZ$ model 
as the scenario in which to analyze how a localized quantum wave packet evolves in time. 
They have shown that besides the quantum mechanical delocalization due to the uncertainty principle, they are in qualitative 
agreement with its classical counterpart. 
Following this objective of deciphering the quantum soliton term, we tackle an alternative problem 
in which we study the time evolution of a one-dimensional topological quantum soliton after a quench.

The study of nonequilibrium phenomena in one-dimensional systems has become a very active area of research in recent years due to
new advances in the experiments wih ultracold atoms in optical lattices, \cite{cold_atoms} 
and the latest studies on thermalization after quantum quenches \cite{out_of_equili, MWNM_09}.
With these ideas in mind, we propose a frustrated $J_1$-$J_2$ spin-$\frac12$ antiferromagnetic Heisenberg chain as a suitable framework in which to
conduct the analysis.
It is know that this model undergoes a phase transition from a quasi-long-range ordered ground state to a product of localized singlets clusters
as a function of the next-nearest neighbor parameter.
Moreover, this Hamiltonian has an exact ground state at the Majumdar-Ghosh (MG) point $J_1 =2J_2$ \cite{Maj_69}, 
and a variational approach describes the elementary excitations adequately \cite{SSBS_81}.
More specifically, 
in this work we propose a fine-tuned magnetic soliton as the initial pattern which will be evolved in time after a quantum quench
of the model parameters. 
In order to create this initial spin-$\frac12$ excitation, 
we choose a one-dimensional chain with spin-phonon coupling as a witness case, which is realized in quasi-one dimensional spin-Peierls systems such as CuGeO$_3$ (Ref.\ \onlinecite{Hase}) 
and TiO$X$ ($X = \text{Cl}, \text{Br}$) \cite{tiocl_seidel}.
An alternative approach would be to prepare the initial excitation and conduct the time evolution in a setup of ultracold atoms in an optical lattice.

Once the initial soliton is prepared, we follow the dynamics of this excitation on the uniform zig-zag Heisenberg chain, driven by a quench of the spin-phonon coupling.
We observe that, in the gapless phase of the model, the initial soliton evolves into two counterpropagating modes, 
indicating a quantum superposition of left- and right- moving components of the original soliton. 
As time evolves, the excitation remains quite stable despite the quantum-mechanical spreading.
On the other hand, for a highly localized soliton generated in the MG point, the excitation shows quick spreading with time evolution.

We select the density matrix renormalization group (DMRG) \cite{White_92} and bosonization \cite{Gia} as the numerical and analytical techniques to conduct our study.
There are plenty of examples in the literature showing that both methods are convenient for giving a reliable description of 
spin chain systems, particularly, when the coupling to the lattice is also considered \cite{DoGa}.
Among different DMRG options \cite{t-DMRG}, we use the algorithm introduced by Manmana {\it et al}, \cite{MWNM_09} 
that allows us to perform non equilibrium simulations for systems with interactions beyond the nearest neighbors.
It is worth mentioning that related studies were done on spin transport, even at
finite temperature \cite{Gobert_05_Langer_09_Jesenko_11_Ganahl_12}.
% \emph{The model and strategy to generate excited states} --
\section{THE MODEL AND STRATEGY TO GENERATE EXCITED STATES}
\begin{figure}[t]
\includegraphics[width=0.47\textwidth]{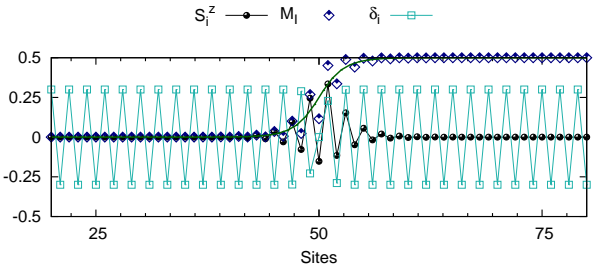}
\caption{(Color online)
Collective excitation obtained by DMRG in a finite lattice of $N_s=99$ sites, in
the absence of frustration ($\beta=0$). The soliton, represented by $\langle S_i^z\rangle$ (circles), was 
tuned by selecting $\delta_i$ in a given configuration (squares). The parameters are $\xi=1, \delta_0=0.3$.
We also show the cumulative magnetization up to a given site (diamonds). The bosonization
result for this magnitude is represented by the solid line.}
\label{fig1}
\end{figure}

In order to generate a soliton-like topological excitation, we introduce a one dimensional antiferromagnetic Heisenberg
Hamiltonian with first and second neighbor interactions and spin-lattice coupling, which reads
\begin{align}
\label{hspindelta}
H&=\sum_i \left[1+ \delta_i\right] \mathbf{S}_{i} \cdot \mathbf{S}_{i+1} + \beta \mathbf{S}_{i-1} \cdot \mathbf{S}_{i+1}.
\end{align}
$\mathbf{S}_i$ is a spin-1/2 operator for the $i$th lattice site, and we have set $J_1=1$ as
the energy scale, such that $\beta=J_2/J_1$ is the second neighbor exchange coupling; $\delta_i=\lambda (u_{i+1}-u_{i})/J_1$ is the
dimensionless bond length variation, where $u_{i}$ are the displacements of the magnetic ions from their equilibrium positions, and
$\lambda$ is the spin-lattice coupling parameter. 
Although frustration is not necessary to present our ideas, we introduce it in order to compare it with the well-known MG 
limit \cite{Maj_69} to achieve a better understanding of the process.
So, with the intention of promoting a sort of quantum topological soliton, we break the lattice symmetry to manipulate
the nearest neighbor magnetic interaction.
From the studies of models with spin-phonon coupling it is known that, when phonons are treated adiabatically, the stable
pattern for $\delta_i$ in the $S^z=1/2$ sector is the one called the static lattice soliton, given by $\delta_i=(-1)^i\delta_0 \tanh(\frac{i-i_0}{\xi})$.
The domain wall is centered at site $i_0$ and its width is given by $\xi$, which produces an interpolation between two dimerized patterns.
The  $S^z=\frac{1}{2}$ magnetic soliton generated by this lattice arrangement  is also centered at site $i_0$, 
and the number of sites involved within the wall is controlled by the parameters $\xi$ and $\delta_0$.
In Fig.\ \ref{fig1} we show the lattice configuration generated with the set of parameters $\{\beta=0, \delta_0=0.3,\xi=1\}$, together with the
associated spin pattern given by $\langle S^z_i\rangle$,
which was obtained from Eq.\ (\ref{hspindelta}) using DMRG in a lattice of size $N_s=99$, where the odd
number of sites is to set the total magnetization to $\frac{1}{2}$. We use open boundary conditions keeping $m=300$
states, enough to assure the accuracy, with a truncation error of order $O(10^{-9})$ in the worst case.
We see that most of the spins arrange in localized singlet clusters, producing zero local magnetization. However, in the center of the
chain where the bonds interpolate between the two possible dimerized states, $\langle S^z_i\rangle \neq 0$. We also show the cumulative
magnetization up to a given site $I$, which is defined as $M_I=\sum_{i=1}^{I} \langle S^z_i\rangle$.
A clear solitonic profile can be observed in this quantity.
Let us now resort to the bosonization technique to analyze the $M_I$ parameter.
In this representation, the $z$-component of the spin is connected to a bosonic field $\phi(x)$ by \cite{Gia} 
\begin{equation}
S^z_i=\frac{1}{2 \pi}\partial_x \phi(x=ia) + \frac{(-1)^i}{\pi \alpha}\cos[2 \phi(x=ia)],
\end{equation}
where $a$ is the lattice constant and $\alpha$ is a short range cutoff in the bosonization procedure. The quantity $M_{I}$ is obtained by
integration of the previous equation up to a point $X$.
Being oscillatory at the lattice level, the last term vanishes. In the  $S^z=\frac{1}{2}$ subspace the field goes from
$\phi(-\infty)=-\frac{\pi}{2}$ to $\phi(\infty)=\frac{\pi}{2}$. Therefore, the expression for $M_I$ is
\begin{equation}
M_I=\frac{1}{2\pi}\langle \phi(X=Ia) \rangle + \frac{1}{4}.
\end{equation}
As the bosonized version of the Hamiltonian (\ref{hspindelta}) is not exactly solvable for general displacements $\delta_i$, we resort to
a semiclassical solution in order to calculate the mean value of the field in the corresponding subspace and compare with the DMRG results.
This solution can be obtained by adding an elastic energy term $\frac{K}{2}\sum_i \delta_i^2$ to the Hamiltonian and treating $\delta_i$
in the adiabatic approximation \cite{NakanoFukuyama,dobryibaceta98}. The classical solutions for the continuous fields are $\delta(x)=\delta_0 \tanh(x/\xi)$
and $\phi(x) = \arcsin\left[\tanh(x/\xi)\right]$.
With the inclusion of an elastic energy term, $\delta_0$ and $\xi$  depend on the values of the
microscopic parameters $K$ and $J_{1,2}$. In the previous DMRG calculation we chose $\delta_i$ with arbitrary $\delta_0$
and $\xi$, i.e. not subject to fulfilling an adiabatic equation. In the spirit of a semi-classical quantization \cite{Rajaramanbook}
we assume the following ansatz for the mean value of the bosonic field: $\langle\phi\rangle \equiv\phi_S(x)=\arcsin\left[\tanh(x/\xi')\right]$,
where the subindex $S$ refers to a solitonic pattern. In Fig.\ \ref{fig1} we show a fitting to the numerical results using the analytical
expression, for which we obtain a good agreement with $\xi^\prime=1.773$.
\begin{figure*}[ht]
\includegraphics[width=0.8\textwidth]{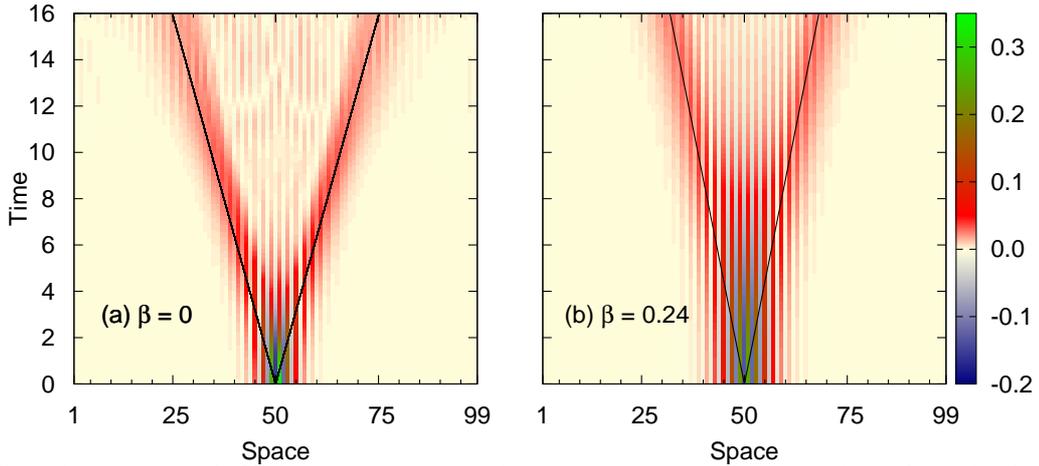}
 \caption{(Color online)
Time evolution of $\langle S_i^z\rangle$ in the collective excitation obtained by DMRG with and without frustration for
the two witness cases, (a) $\beta=0, \delta_0=0.3, \xi=1$ and $\beta=0.24, \delta_0=0.3, \xi=3$. In (a), the black
line corresponds to the slope in the time-space diagram of the spin velocity $v_s=\frac{\pi}{2}$ obtained by the Bethe ansatz
for a homogeneous Heisenberg chain. In (b), the black line shows the spin velocity renormalized by the frustration.
In both cases one observes the splitting of the soliton into two chiral modes.
}
\label{fig2}
\end{figure*}
% \emph{Quench and time evolution of the soliton} --
\section{QUENCH AND TIME EVOLUTION OF THE SOLITON}
We now turn off the spin-lattice coupling and study the dynamics of this magnetic excitation, which will be conducted by a 
homogeneous Heisenberg Hamiltonian $H'\equiv H[\delta_i=0]$.
Once the quantum soliton is constructed, the time-dependent DMRG (t-DMRG) algorithm enables us to evolve in time under this new homogeneous Hamiltonian \cite{warn2}.
As a gapless phase is stable for $0\leq\beta\leq \beta_c\! = \!0.245$, we analyze different tuned solitons in this zone where the bosonization
analysis is valid. In Fig.\ \ref{fig2}, we show the dynamics of two witness cases which are defined by
$\{\beta\!=\!0, \delta_0\!=\!0.3, \xi\!=\!1\}$ and $\{\beta\!=\!0.24, \delta_0\!=\!0.3, \xi\!=\!3\}$.
One can observe that the excitations behave in a subtle way. 
In both cases, the time evolution shows how the original soliton evolves in such a way that, at a
given time, two spin clouds are observed, each of them carrying  $\meanvalue{S^z}\!=\!\frac{1}{4}$.
Remarkably, these right and left modes are quite stable: as observed in Fig.\ \ref{fig2}, these modes barely disperse as time evolves. 
We will return to this interpretation later.

These left and right excitations travel at a velocity which agrees very well with the spin wave velocity of the low energy excitations of the homogeneous chain.
This can be seen in Fig.\ \ref{fig2} by comparing the evolution of the maximum of $\meanvalue{S_i^z}$ with the spin wave velocity
$v_s\!=\!\frac{\pi}{2}$ predicted by the Bethe ansatz for the one dimensional homogeneous Heisenberg chain \cite{Gia}, and the
renormalized $v_s\!=\!\frac{\pi}{2}(1-1.12 \beta)$ calculated in Ref.\ \onlinecite{FG_97} using exact diagonalization in small chains. 

\begin{figure}[b]
\centering
\includegraphics[width=0.45\textwidth]{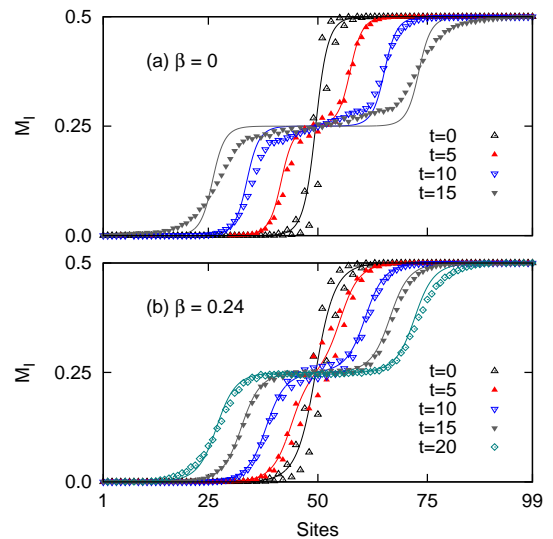}
\caption{(Color online)
Time evolution of the cumulative magnetization  as given by DMRG (symbols) and bosonization (solid lines) in the
(a) non frustrated and  (b) frustrated ($\beta\!=\!0.24$) cases. The observed jump to $M_I=1/4$ as time evolves, signals 
the quantum interference of left and right soliton states.}
\label{fig3}
\end{figure}

Let us analyze the origin of this behavior by studying the time evolution in the bosonization language.
The Hamiltonian $H'$ in the bosonic representation corresponds to a (1+1) free bosonic field theory, up to marginally irrelevant
operators \cite{Gia}. These marginal operators vanish at $\beta\!=\!\beta_c$.
The time evolution of the mean value of $\phi$ is given by
$\langle \phi(x,t) \rangle = \langle e^{-iH't}\phi_R(x) e^{iH't} \rangle +  \langle e^{-iH't}\phi_L(x) e^{iH't}\rangle
= \frac{1}{2}  \left[\phi_S(x-v_st)+\phi_S(x+v_st)\right]$.
To obtain the previous expression, we have taken into account that, in the initial state, the field $\phi$ is split into its left and
right parts $\phi=\frac{1}{2}(\phi_L+\phi_R)$ originating from the left  and right fermions in the bosonization procedure.
For a time-independent state, $\langle \phi_R\rangle = \langle \phi_L\rangle = \frac{\phi_S(x)}{2}$ as in the initial situation discussed above.
On the other hand, the time evolution of $\phi_L$ is independent of that
of $\phi_R$, and it is given by a simple shift in $vt$ to the left and to the right, respectively.
In Fig.\ \ref{fig3} we show the cumulative magnetization for different times copared to the DMRG results. We used  the fitted value to
the static solution for $\xi'$ in $\phi_S$, as shown for the $\beta=0$ case in Fig.\ \ref{fig1}. A similar procedure gives $\xi'=2.781$ for
the $\beta=0.24$ case.
We also fix $v_s$ to the spin-wave velocity of the homogeneous chains as previously discussed.
We observe a fairly good comparison between the DMRG and bosonization results, especially when $\beta$ is near the critical value $\beta_c$.
This is due to the reduction of the finite-size effect in the latter case: as the evolution of the resulting excitations is slower as $\beta$ increases,
they do not reach the edge of the chain for the last time, $t=20$, obtained in our calculation. The improved fitting between both results could also
be due to the fact that the marginal irrelevant term neglected in the bosonized Hamiltonian could play some role in the short-distance correlations
which, nonetheless, vanishes at $\beta_c$.
As in the numerical results, we see that the original soliton does not propagate in a particular direction but the state splits into two 
counterpropagating modes as 
a consequence of the chiral symmetry of the Hamiltonian. As this pattern arises as a sum of two topological protected
excitations, one on the right sector and one on the left sector of the theory, we can assume that the global excitation is also protected and 
remains stable with the time evolution.
The jump in $M_I$ indicates that each chiral mode carries $\meanvalue{S^z}\!=\!\frac{1}{4}$.
This can be interpreted as a superposition $(\ket{L} + \ket{R})/\sqrt{2}$ of two $S^z=1/2$ quantum soliton states, 
such that $\ket{L}$ $(\ket{R})$ is a state propagating to the left (right).
The system retains the memory of the initial state, but the original soliton state evolves into two chiral states as the time 
evolution is carried out with a Hamiltonian whose left and right modes are independent.
However, the elementary excitations of the uniform Heisenberg model are not these type of solitons, notably the dynamical states preserve
coherence and disperse quite slowly.

It is worth remarking that working with an odd number of sites is not a necessary condition for these results. 
An even number of sites will sustain a pair of soliton and antisoliton excitations. For long enough separation between them in order to prevent interaction 
effects, the conclusion remains the same, so that each soliton will split into two left and right modes as time evolves.

One wonders if in the limiting case of a soliton of zero width,
i.e., at the MG point ($\beta\!=\!1/2$), such mechanism could come into effect.
% \emph{Soliton of zero width: a free spin} --
\section{SOLITON OF ZERO WIDTH: A FREE SPIN}
The MG point is a good scenario to create such localized excitation, and as it has an exact eigensolution built of singlets,
we analyze the case in order to compare our numerical calculations with the variational approach of Ref.\ \onlinecite{SSBS_81}.
To recreate this situation we use  $\delta_i=0$ and $\beta\!=\!1/2$ in Eq. (\ref{hspindelta}), and set to zero
the exchange parameters connecting to site $i=(N_s+1)/2$ as well. Thereby, 
we establish a free spin at the center of the lattice, separating
two MG domains $\ket{\spinon}$ with singlets $\ket{\dimer}\!=\!\frac{1}{\sqrt{2}}[\ket{\ups\downs}-\ket{\downs\ups}]$.
Since the physics we want to describe is localized, we choose a lattice of $N_s\!=\!41$ in the  $S^z\!=\!\frac12$ subspace without worrying
about the edge effects.

\begin{figure}[t]
\includegraphics[width=0.4\textwidth]{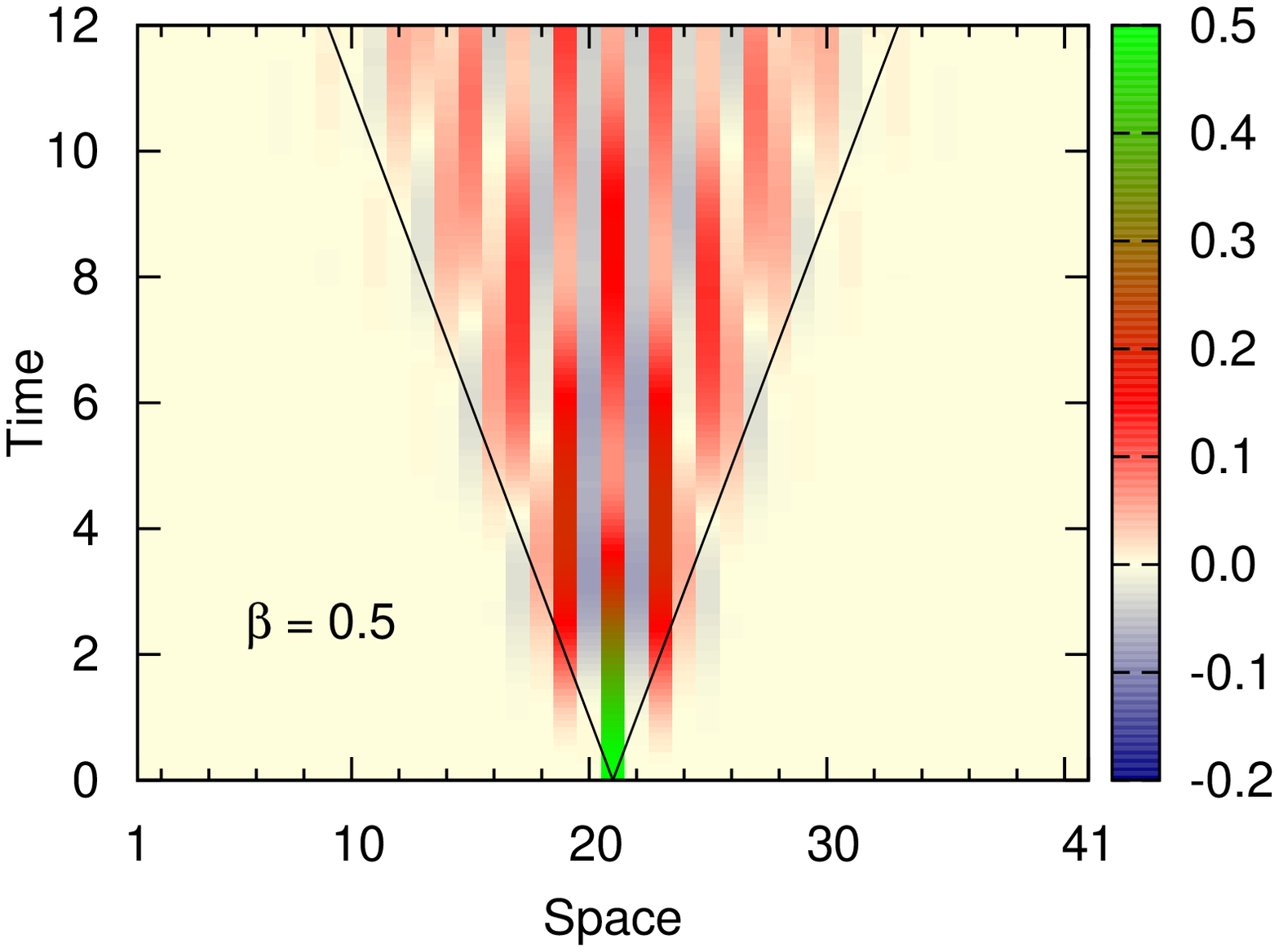}
\vspace{-0.5cm}
\caption {(Color online)
Time evolution of $\meanvalue{S_i^z}$ for the free spin on a lattice of 41 sites at the MG point $\beta\!=\!1/2$.
The solid lines correspond to the maximum group velocity according to the
variational dispersion relation obtained from Ref.\ \onlinecite{SSBS_81}.
}
\label{fig4}
\end{figure}

In Fig.\ \ref{fig4} we show the time evolution of $\langle S_i^z\rangle$ once the uniform zigzag Heisenberg model is restored, from which 
we can appreciate two things. First, we have  
appropriate agreement with the variational approach \cite{SSBS_81}, where elementary excitations are described
by $\omega(k)\!=\!2\beta[\frac{5}{4}\!+\!cos(2k)]$.
The slope of the solid lines in Fig.\ \ref{fig4} agrees with the maximum group velocity that results from the variational dispersion relation.
We are not aware of other numerical calculations showing this in the literature.
Secondly, we have qualitative agreement with the effective Hamiltonian description proposed in Ref.\ \onlinecite{ALGR_12} for the dynamics of a free
spin hopping between next nearest neighbors. It is easy to appreciate that the initial free spin reduces its module at the center
site with the time evolution, transferring its spin component to the next nearest neighbor. 
The spin uses the mechanism of exchange mediated by
$J_2$ to avoid breaking the dimers, which would result in a loss of magnetic energy. 
Different from the mechanism previously described in which we found a separation into chiral modes, here the time evolution of $\langle S_i^z\rangle$ shows 
the expected dispersion of the original individual excitation moving in a singlet sea. 
The initial excitation cannot remain highly localized because of the uncertainty principle. As the quasiparticle is localized in real space, it involves
a broad range of momenta in the reciprocal space, in which case the excitation spreads very quickly.
% \emph{Possible experimental realizations} --
\section{POSSIBLE EXPERIMENTAL REALIZATIONS}
Two experimental realizations of the previous mechanism are envisaged. One is on spin-Peierls materials 
such as CuGeO$_3$ (Ref.\ \onlinecite{Hase}) or TiO$X$ ($X = \text{Cl}, \text{Br}$) \cite{tiocl_seidel}.
At low temperatures, below a critical value $T_{\text{SP}}$, these materials undergo a magneto structural transition in which the lattice
dimerizes and a spin gap opens in the magnetic spectrum. Applying a magnetic field larger than the critical one $H_{\text{c}}$, a soliton lattice is generated 
which has been characterized by x-ray scattering measurements \cite{Keimer}.
For not too strong a magnetic field above $H_{\text{c}}$, the magnetic chains realize our initial state because the solitons are far apart.
Then, the field should be turned off and the temperature raised above $T_{\text{SP}}$. The system should now be in the uniform phase and the magnetism
should be described by a homogeneous Heisenberg Hamiltonian, which will conduct the time evolution as well.
Another possible realization is in recent experiments of ultracold atoms trapped in a one-dimensional optical lattice.
Recently, the bosonic repulsive Hubbard model has been successfully accomplished \cite{Takeshi_13} to study the dynamics of spin excitations
on a ferromagnetic background.
On the other hand, the fermionic version has been realized \cite{Esslinger}, and once the limitations related to the
temperature are overcome, it is expected that the antiferromagnetic Heisenberg model can be simulated by tuning the Feshbach resonances for large on-site repulsion.
%
%\emph{Conclusions} --
\section{CONCLUSIONS}
We described a mechanism based on the preparation of a soliton, which after a quench of the interaction drives the system to
the formation of a state composed of two counterpropagating components as times evolves, producing two-well defined spin clouds in the chain, 
each of them with total magnetization $\meanvalue{S^z}\!=\!1/4$.
This state can be interpreted as a quantum superposition of left and right moving $S^z=1/2$ quantum soliton states that
are protected by the chiral separation of the Hamiltonian. Remarkably, they remain stable as time evolves, showing a very slow
dispersion.
We propose a recipe to obtain these states, consisting of an initial preparation of a topological protected excitation by a modulation of the exchange
couplings, followed by a quench of the interactions, generating new left and right solitonic states that will evolve in time through a Hamiltonian whose
left and right modes are independent.
On the other hand, the situation for the MG point is different. As the initial excitation is local in real space, 
its distribution in momentum space is very broad, leading to a rapid dispersion of the excitation as time evolves.
The most natural candidates to observe those phenomena would be spin-Peierls materials and ultracold atoms in optical lattices.

\section*{ACKNOWLEDGMENTS}
We thank Armando Aligia, Daniel Cabra, Anibal Iucci, Luis Manuel, and Adolfo Trumper for useful discussions.
D.\ M.\ acknowledges the hospitality of the Dahlem Center.
This work was partially supported by PIP CONICET Grant 0392, Grant DMR-1108285 (Ohio), and NSF-PIRE Grant 0730257.


\begin{thebibliography}{99}
\bibitem{warn2} We applied a time step  $\Delta t = 0.1$ and kept 100 Lanczos vectors in the
Krylov-space representation, resulting in a discarded weight of order $O(10^{-7})$ at the end
of the time evolution. The error bars result to be smaller than the size of the symbols.
\end{thebibliography}
\end{document}